
\input phyzzx
\PHYSREV
\hoffset=0.3in
\voffset=-1pt
\baselineskip = 14pt \lineskiplimit = 1pt
\frontpagetrue
\rightline {Cincinnati preprint October.1994}
\medskip
\titlestyle{\seventeenrm Renormalizability and Quantum Stability
of the Phase Transition in Rigid String Coupled to
Kalb-Ramond Fields II}
\vskip.5in
\medskip
\centerline {\caps M. Awada\footnote*{\rm E-Mail address:
moustafa@physunc.phy.uc.edu}}
\centerline {Physics Department}
\centerline {\it University of Cincinnati,Cincinnati, OH-45221}
\bigskip
\centerline {\bf Abstract}
\bigskip
Recently we have shown that a phase transition occurs
in the leading approximation of the large N limit in rigid strings
coupled to long range  Kalb-Ramond interactions.  The disordered
phase is
essentially the Nambu-Goto-Polyakov string theory while the ordered
phase
is a new theory.  In this part II letter we study the first
sub-leading
quantum corrections we started in I.  We derive the renormalized
mass gap equation and obtain the renormalized critical line of the
interacting theory.  Our main and final result is that the
phase transition does indeed survive quantum fluctuations.

\eject

Polyakov [1] has argued that the string theory appropriate to QCD
should
be one with long range correlations of the unit normal. The
Nambu-Goto
(NG) string theory is not the correct candidate for large N QCD as it
disagrees with it at short distances.  The NG string does not give
rise to the parton-like behaviour observed in deep inelastic
scattering
at very high energies.  The observed scattering amplitudes have a
power
fall-off behaviour contrary to the exponential fall-off behaviour of
the NG string scattering amplitudes at short distances.  The absence
of
scale and the power law behaviour at short distances suggest that
the QCD string must have long range order at very high energies.
Pursuing this end, Polyakov considered modifying the Nambu action
by the renormalizable scale invariant curvature squared term
(rigid strings).  The theory closely resembles the two dimensional
sigma model where the unit normals correspond to the sigma fields.
In the large N approximation, there is no phase transition.
Polyakov suggested adding a topological term to produce a phase
transition to a region of long range order.  In [2],
we coupled the rigid strings instead to long range Kalb-Ramond
fields.
Since spin systems in two dimensions may exhibit a phase
transition with the inclusion of long range interactions, it is
natural to conjecture likewise for rigid strings
with long range Kalb-Ramond fields.  Indeed  we proved that there is
a phase transition to a region of long range order in
the large N approximation. Such a theory may therefore be
relevant to QCD.

Our proof in [2] of the phase transition was based on
the leading order approximation in large N, where N is the space-time
dimensions.  Even though, the large N limit is a successful
approximation
for non-linear sigma models, and some spin systems it can
sometimes lead to an incorrect conclusion.  The leading order
of the large N approximation is mean field theory which can give
incorrect predictions in lower dimensions.  For example mean field
theory incorrectly gives a phase transition in the one
dimensional Ising model. This discrepancy is
resolved by carefully examining the sub-leading quantum corrections
(loops) where one shows that such quantum corrections in fact destroy
the phase transition. Therefore it is crucial to examine
the quantum loop corrections to the mass gap, and the critical line
of
our model of rigid string coupled to long range Kalb-Ramond fields.

In this letter we will generalize the analysis we started in I and
prove that the phase transition in our model
survives quantum fluctuations and that the quantum loop
corrections to the sub-leading order lead to mass and wave function
renormalizations.  Similar prove was given for the phase transition
in
rigid QED [3]

 The gauge fixed action of the rigid string [2] coupled to the rank
two antisymmetric Kalb-Ramond tensor field $\phi$ is [4]:
$$S_{gauge-fixed} = \mu_{0}\int d^2\xi\rho + {1\over 2t_{0}}\int
d^2\xi
[\rho^{-1}(\partial^{2}x)^{2} + \lambda^{ab}(\partial_{a}x
\partial_{b}x - \rho\delta_{ab})] + S_{K-R}\eqno{(1 a)}$$
where
$$S_{K-R}= e_{0}\int d^2\xi \epsilon^{ab}\partial_{a}x^{\mu}
\partial_{b}x^{\nu}\phi_{\mu\nu} + {1\over 12}\int d^4x
F_{\mu\nu\rho}F^{\mu\nu\rho}\  .\eqno{(1 b)}$$
where $e_{0}$ is a coupling constant of dimension $length^{-1}$,
$t_{0}$ is the bare curvature coupling constant which is
dimensionless
and F is the Abelian field strength of $\phi$.  The integration of
the
$\phi$ field is Gaussian.  We obtain the following interacting
long range Coulomb-like term that modifies the rigid string:
$${1\over 2t_{0}}\int\int d^2\xi d^2\xi' \sigma^{\mu\nu}(\xi)
\sigma_{\mu\nu}(\xi')V(|x-x'|, a)\eqno{(1 c)}$$
where V is the  analogue of the long range Coulomb potential:
$$ V(|x-x'|,a)= {2g_{0}\over \pi}{1\over |x(\xi)-x(\xi')|^{2}
+a^{2}\rho}\   .\eqno{(1 d)}$$
where $\sigma^{\mu\nu}(\xi)= \epsilon^{ab}\partial_{a}x^{\mu}
\partial_{b}x^{\nu}$.
We have introduced the cut-off "a" to avoid the singularity at
$\xi=\xi'$ and define $ g_{0}=t_{0}\alpha_{Coulomb}=t_{0}
{e_{0}^{2}\over 4\pi}$ which has dimension of $length^{-2}$.
The partition function is
$$ Z =\int D\lambda D\rho Dx exp(-S_{eff})\  .\eqno{(2)}$$
where the effective action $S_{eff}$ is (1a) and (1c).

The long range KR interactions are non-local and impossible to
integrate.  Therefore we consider
$$x^{\nu}(\xi) = x^{\nu}_{0}(\xi) + x^{\nu}_{1}(\xi)$$
and expand the Coulomb term (1c,d) to quadratic order in
$x^{\nu}_{1}(\xi)$ about the background straight line $x_{0}$.
The x-integration is now Gaussian and to the leading D approximation
we obtain the new effective action $S_{eff}$:
$${S_{0}}_{eff} = {1\over 2t_{0}}[\int d^2\xi (\lambda^{ab}
(-\rho\delta_{ab}) + 2t_{0}\mu_{0}\rho) + t_{0}Dtrln A]\eqno{(3)}$$
where A is the operator
$$ A = \partial^{2}\rho^{-1}\partial^{2} -\partial_{a}
\lambda^{ab}\partial_{b} + V(\xi,\xi')\  .\eqno{(4)}$$
In the large D limit the stationary point equations resulting
from varying $\lambda$ and $\rho$ respectively are:
$$\rho={t_{0}D\over 2}trG\eqno {(5 a)}$$
$$2t_{0}\mu_{0}-\lambda^{ab}\delta_{ab}=t_{0}Dtr(\rho^{-2}
(-\partial^{2}G))\eqno {(5 b)}$$
where the world sheet Green's function is defined by:
$$ G(\xi,\xi') = <\xi|(-\partial^{2})A^{-1}|\xi'>\eqno{(6)}$$
The stationary points are:
$$ \rho(\xi) = \rho^{*},~~~~~~~~\lambda^{ab}=\lambda^*\delta^{ab}
\eqno{(7)}$$
where $\rho^{*}$ and $\lambda^*$ are constants.  Using the stationary
solutions (7) the operator A is given by:
$$ trln A = A_{\Sigma}\int{d^2p\over (2\pi)^2}ln[p^{4}
+ p^{2}m^{2} + p^{2}V_{0}(p) + V_{1}(p)]\eqno{(8)}$$
where $A_{\Sigma}$ is the area of the surface and
$$ V_{0}(p) = {4g_{0}\rho^{*}\over \pi}\int d^2\xi
{e^{ip.\xi}\over \xi^{2}+a^{2}}= 8g_{0}\rho^{*} K_{0}(a|p|)\eqno{(9
a)}$$
$$V_{1}(p)={8g_{0}\rho^{*}\over \pi}\int d^2\xi
{[e^{ip.\xi}-1]\over (\xi^{2}+a^{2})^{2}}= {8g_{0}\rho^{*} \over a^2}
(a|p|K_{1}(a|p|)-1)\eqno{(9 b)}$$
where $K_{n}(z), n=0,1,..$ is the Bessel function of the third
kind and
$$K_{1}(z) := -{d\over dz}K_{0}(z)\  .\eqno{(10)}$$
Thus (5a) becomes the mass gap equation [2]:
$$ 1 ={Dt_{0}\over 2}\int{d^2p\over (2\pi)^2}
{p^{2}\over p^{2}(p^{2}+m_{0}^{2}) + p^{2}V_{0}(p)
+ V_{1}(p)}\eqno{(11)}$$
where we have defined the bare mass:
$$m_{0}^{2}=\rho^*\lambda^*\eqno{(12)}$$
and it is  associated with the propagator:
$$ <\partial_{a} x^{\mu}(p)\partial_{a} x^{\nu}(-p)> =
{Dt_{0}\over 2}{\delta^{{\mu}{\nu}}\over p^2+m_{0}^2 +
{\cal V}(p^2)}\eqno{(13)}$$
where
$${\cal V}(p^2) =  V_{0}(a^2p^2) + {V_{1}(a^2p^2)\over p^2}
\  .\eqno{(14)}$$
On the other hand eq(5b) yields the string tension
renormalization condition:
$$\mu_{0}= {D\over 8\pi}{\Lambda^{2}\over \rho^{*}}
- {D\over 2\rho^{*}}\int{d^2p\over (2\pi)^2}
{{\cal V}(p^{2})\over p^{2}+m_{0}^{2} + {\cal V}(p^2)} \eqno{(15)}$$
where $\Lambda={1\over a}$ is an U.V. cut-off.
To obtain a  non-zero phase transition temperature the mass gap
equation must be infra-red finite for $m_{0}^2=0$.  Therefore
without the K-R long range interactions ($g_{0}=0$) the theory exists
only in the disordered phase $t_{0}>t_{c}$ and the U.V stable fixed
point is $t_{c}=0$.  In this case the beta function of the
free rigid string theory is asymptotically free, indicating the
absence of the extrinsic curvature term at large distance scales.
In contrast to the naive classical limit the theory is therefore
well behaved and free of ghosts.  As we have shown in I this property
prevails in the sub-leading quantum corrections to the mass gap
equation.
The ordered and disordered phases are separated by the
critical line defined by eq.(11) at $m_{0}^2=0$:
$$ 1 ={Dt_{0}\over 4\pi}\int_{0}^{\epsilon} dy{y^{3}\over y^{4}+
\kappa_{0}(y^{2}K_{0}(y)+yK_{1}(y)-1)}\eqno{(16)}$$
where $\kappa_{0}=8g_{0}\rho^{*} a^{2}$ is a dimensionless coupling
constant. We have made the change of
variable y=ap and introduced the U.V cut-off
$\Lambda$, and $\epsilon={\Lambda\over \Lambda_{0}}$
where $\Lambda_{0} ={1\over a}$.  It is straightforward to prove
that there exist an $\kappa^{*}$ (c.f. Fig.1)
for which any choice of $\epsilon$ leads to phase transition as long
as $\kappa_{0}<\kappa^{*}$.  We set $\epsilon = 1$. It is remarkable
that eq.(16) is finite except at $\kappa_{0}=0$
(absence of K-R interactions). In fact after tedious
calculations one can prove that
$$ lim_{y\rightarrow 0}
[{y^{3}\over y^{4}+\kappa_{0}(y^{2}K_{0}(y)+yK_{1}(y)-1)}]=0$$
thus having an infra-red finiteness.

The dimensionless coupling constant $\kappa_{0}=8g_{0}\rho^{*} a^{2}$
has a natural interpretation.  From the string renormalization
condition (15) one obtains the critical value of the string tension
$\mu_{c}$ to be
$$\mu_{c} = {D\over 4\pi a^2 \rho^{*}}\int_{0}^{1} dy{y^{5}\over
y^{4}+
\kappa_{0}(y^{2}K_{0}(y)+yK_{1}(y)-1)}$$
i.e
$$\mu_{c} :={D\over 8\pi a^2 \rho^{*}}f(\kappa_{0})$$
where f is a positive function of $\kappa_{0}$.  Combing the above
result
with the definition of $\kappa_{0}$ we finally obtain:
$$\kappa_{0} = h({D g_{0}\over \pi \mu_{c}})$$
as a positive function of the ratio of the Kalb-Ramond coupling
constant
to the critical string tension.

The critical curve distinguishing the two phases in
the $(t_{0},\kappa_{0})$ plane is shown in Fig.1.  The order
parameter of the theory is the mass gap equation
(15) where $m_{0}^2$ is the parameter that distinguishes
that two phases.  In the disordered phase $ m_{0}^2 >0$ and
$<\partial x_{c}^{0}> = 0$, while in the ordered phase it is
straightforward
to show that $m_{0}^2 =0$ and the order parameter is the analogue of
the
classical magnetization, namely $<(\partial x_{c}^{0})^2>$, thus the
mass
gap is
$$<(\partial x_{c}^{0})^2> = 1 - {Dt_{0}\over 2}\int{d^2p\over
(2\pi)^2}
{1\over p^{2} + {\cal V}(p^2)}.$$
In the disordered phase the coupling constants $t_{0}$ and
$\kappa_{0}$ are
completely fixed by dimensional transmutation in terms of the cut-off
$\Lambda^2$ and $m_{0}^2$. Thus, they cannot be fine tuned.
This is an important property that is vital in proving the absence
of ghosts in our model [5].
\bigskip
{\bf II-The Loop Corrected Mass Gap Equation}

In mean field theory i.e leading order in ${1\over D}$,
the relevant propagator is equation (13).  In the sub-leading
correction to mean field theory, the quantum fluctuation
imply a new term corresponding to the self-energy of the
$\partial_{a} x^{\mu}$-field
$$ <\partial_{a} x^{\mu}(p)\partial_{a} x^{\nu}(-p)> =
{Dt_{0}\over 2}{\delta^{{\mu}{\nu}}\over (p^2+m_{0}^2
+ {\cal V}(p^2) + {1\over D}\Sigma(p))}\  .\eqno{(17)}$$
The new contribution $\Sigma (p)$ arises from fluctuations of the
Lagrange multipliers $\lambda_{ab}$ and $\rho$ where the fluctuations
$\sigma_{ab}$ and $\eta$ are defined by:
$$ \lambda_{ab} = \lambda^{*}\delta_{ab} +
i{1\over \sqrt{(D/2)}}\sigma_{ab}$$
$$\rho = \rho^{*}(1 + i{\rho^{*}\over \sqrt{(D/2)}}\eta)\
.\eqno{(18)}$$
Expanding the effective action (3) in powers of $\sigma_{ab}$ and
$\eta$,
it is straightforward to extract the $\sigma_{ab}$ and
$\eta$ propagators (Fig (2)):
$$ \Pi_{ab|cd}(p) = {1\over 2}(\delta_{ac}\delta_{bd} +
\delta_{ad}\delta_{bc})\pi(p^2)\eqno{(19 a)}$$
$$\pi(p^2) =
\int {d^{2}k\over (2\pi)^{2}} {1\over (k^2 + m_{0}^{2} + {\cal
V}(k^2))
((p + k)^2 + m_{0}^{2} + {\cal V}((p+k)^2))}\eqno{(19 b)}$$
$${\tilde \pi}(p^2) =
2\int {d^{2}k\over (2\pi)^{2}}{(k.(k+p))^2\over (k^2 + m_{0}^{2} +
{\cal V}(k^2))((p + k)^2 + m_{0}^{2} + {\cal V}((p+k)^2))}
\  .\eqno{(19 c)}$$
It is obvious from (19c) that the $\eta$ propagator has quadratic and
logarthmic divergences and therefore needs regularization before
attempting to analyze the self energy.  These type of divergences
were
exactly calculated in I, and therefore can be subtracted in a
similar fashion here.  The only exception is that we cannot find an
exact analytical answer for both $\pi$ and the regularized $\pi^{*}$
because of the complexity and the non-polynomial structure of the
potential ${\cal V}(p^2)$.  Fortunately, we do not need the exact
analytical forms of the $\pi$'s because all we are interested in is
the ultra-violet behaviour of the self energy $\Sigma$ from which we
can
extract the divergent contributions.  For this we only need the
asymptotic forms of the propagators and they are:
$$\pi_{asy}(p^2) = {1\over 2\pi (p^2 +{\cal V}(p^2))}
log{p^2\over m_{0}^2}\eqno{(20 a)}$$
$${\pi^{*}}_{asy}(p^2) = {p^2 +{\cal V}(p^2)\over 4\pi}
log{p^2\over m_{0}^2}\  .\eqno{(20 b)}$$

The self energy $\Sigma$ can be computed from the diagrams of
Fig(3).  These diagrams are of order ${1\over D}$ and represent
the quantum fluctuations:
$$\Sigma(p) = \int{d^2k\over (2\pi)^2}
{\pi^{-1}(k^2)\over ((p + k)^2 + m_{0}^2 +{\cal V}((p+k)^2))}$$
$$+ {1\over 2}\int{d^2k\over (2\pi)^2}
{{\pi^*}^{-1}(k^2)(k.(k+p))^2\over ((p + k)^2 + m_{0}^2 +
{\cal V}((p+k)^2))}$$
$$ - \int {d^2k\over (2\pi)^2} \int{d^2q\over 2\pi} {\pi^{-1}(0)
\over (q^2 + m_{0}^2 +{\cal V}(q^2))^{2}}
{\pi^{-1}(k^2)\over ((q + k)^2 + m_{0}^2 +{\cal V}((q+k)^2))}$$
$$ -{1\over 2}\int {d^2k\over (2\pi)^2} \int{d^2q\over 2\pi}
{\pi^{-1}(0)
\over (q^2 + m_{0}^2 +{\cal V}(q^2))^{2}}
{{\pi^*}^{-1}(k^2)(k.(k+q))^2\over ((q + k)^2 + m_{0}^2 +
{\cal V}((q+k)^2))}\  .\eqno{(21)}$$
A Taylor expansion of the self energy about zero momentum leads
to mass and wave function renormalizations and a remaining piece
${\tilde \Sigma}$ which must be finite for the theory to be
renormalizable.  The propagator now reads:
$${Z\over (p^2 + m^2 + {\cal V}(p^2) + {1\over D}
{\tilde \Sigma_{finite}(p)})}\  .\eqno{(22)}$$
where
$$ Z = 1 - {1\over D} \Sigma'(0)\eqno{(23 a )}$$
is the wave function renormalization,
$$m^2 =m_{0}^2 + {1\over D}(\Sigma(0) -m_{0}^2
\Sigma'(0))\eqno{(23 b )}$$
is mass renormalization and
$$ \kappa = Z\kappa_{0}\eqno{(23 c)}$$
is the K-R coupling renormalization. The renormalized mass gap
equation is:
$$ 1 = {Dt\over 2}\int {d^2p\over (2\pi)^2}
{1\over (p^2 + m^2 + {\cal V}(p^2) + {1\over D}
{\tilde \Sigma}_{finite}(p))}\  .\eqno{(24)}$$

To examine whether there is still a phase transition
i.e the infra-red finiteness of the renormalized mass gap equation
eq.(24) at $m^2 = 0$, we must examine the renormalizability of the
theory.   We have seen in the examples of rigid QED [3] and the free
rigid string in part I that all typical divergences come from
$\Sigma(0)$
and $\Sigma'(0)$  which can be absorbed in mass and wave function
renormalizations.  In order to calculate the finite regularized self
energy
we need to simplify further Eq.(21).  Using (19) we can replace
zero-momentum insertions of the $\sigma$ and $\rho$ fields and
rewrite (21) as:
$$\Sigma(p) = \Sigma_{1}(p) + {1\over 2}I_{1}\int{d^2k\over (2\pi)^2}
{\pi^*}^{-1}(k^2) - {1\over 2}\int{d^2k\over (2\pi)^2} k^2
{\pi^*}^{-1}(k^2)
 \eqno{(25)}$$
$$+\int{d^2k\over (2\pi)^2}
{\pi^{-1}(k^2)\over ((p + k)^2 + m_{0}^2 +{\cal V}((p+k)^2))} +
{1\over 2}\pi(0)^{-1}
\int{d^2k\over (2\pi)^2}\pi^{-1}(k^2){\partial\over
m_{0}^2}\pi(k^2)$$
$$ +{1\over 2}\int{d^2k\over (2\pi)^2}
{{\pi^*}^{-1}(k^2)(k.(k+p))^2\over ((p + k)^2 + m_{0}^2 +
{\cal V}((p+k)^2))} + {1\over 4}\pi(0)^{-1}\int{d^2k\over (2\pi)^2}
{\pi^*}^{-1}(k^2){\partial\over m_{0}^2}{\pi^*}(k^2)$$
where $\Sigma_{1}(p)$ is a finite piece of the self energy and
$$I_{1}({\Lambda^2\over m_{0}^2},\kappa_{0})=
\pi(0)^{-1}\int {d^{2}q\over (2\pi)^{2}} {1\over (q^2 + m_{0}^{2} +
{\cal V}(q^2))} = {2\pi(0)^{-1}\over Dt_{0}}\  .\eqno{(26)}$$
where we have used the mass gap eq. (11).  In deriving (25) we have
used the facts that $lim_{p \rightarrow \infty} {{\cal V}(p^2)\over
p^{n}}
=0,~~ n\geq -1$ which follows from the asymptotic properties of the
Bessel
functions of the third kind.

As in part I we can now regularize (25) by applying the SM
regularization
scheme [4] where one subtracts the
highest powers of the integration variable appearing in the Taylor
expansion of the integrands in (25).  Again the divergences in (25)
occur
only at the zeroth and first order expansions.  Using the asymptotic
limits
of the propagators (19) we finally obtain the following finite
regularized self energy:
$$\Sigma_{finite}(p) = \Sigma(p) + I_{1}(1- {1\over 2}I_{0})
 -{m_{0}^2\over 4} I_{0} - {1\over 4}(p^2 + m_{0}^2) I_{0}\eqno{(27
a)}$$
where
$$I_{0}({\Lambda^2\over m_{0}^2},\kappa_{0}) =
\int{d^2k\over (2\pi)^2}{4\pi\over (k^2 +{\cal V}(k^2))
log{k^2\over m_{0}^2}} \  .\eqno{(27 b)}$$
The results (26) and (27) generalize equation (23) in part I for the
free
rigid string to the interacting rigid string with long range K-R
fields.

 We can now read from (27) and (23c) the mass, wave function, and K-R
renormalizations:
$$ m^2 = m_{0}^2[1 - {1\over D}({I_{1}\over m_{0}^2}
(1- {1\over 2}I_{0}) -{1\over 4} I_{0})]\eqno{(28 a)}$$
$$ Z = 1 - {1\over 4D}I_{0}\  .\eqno{(28 b)}$$

The beta function can now be obtained as in part I by holding
$m^2(\Lambda,m_{0},t)$ fixed:
$$\beta(t) = -({\partial log m^2\over \partial t})^{-1}\
.\eqno{(29)}$$
Having shown that the theory is renormalizable we can now give the
renormalized critical line defined by (24) at $m^2 =0$
$$1 = {Dt\over 2}\int {d^2p\over (2\pi)^2}
{1\over (p^2 + {\cal V}(p^2) + {1\over D}
{\tilde \Sigma}_{finite}(p))}\eqno{(30)}$$
which is certainly is Infra-red finite since by its definition
${\tilde \Sigma_{finite}(0)}=0$
and (30) then has the exact behaviour as (16) at y=0.  Furthermore,
as long
as $\kappa<\kappa^{*}$ we have shown that the critical line (16) has
no real
poles. i.e $(p^2+{\cal V}(p^2)) > 0$. The presence of
${\tilde \Sigma}_{finite}$ will not
affect the above conclusion to any sub-leading finite order in
perturbation
theory because in the sub-leading ${1\over D}$ order the critical
line is:
$$ 1 = {Dt\over 2}\int{d^2p\over (2\pi)^2}({1\over (p^2+
{\cal V}(p^2))} - {1\over (p^2+
{\cal V}(p^2))}{1\over D}{\tilde \Sigma}_{finite}(p)
{1\over (p^2+{\cal V}(p^2))})\  .\eqno{(31)}$$
As in any quantum field theory the location of the poles in the
presence
of ${\tilde \Sigma}_{finite}(p)$ is a non-perturbative issue and
requires the
form of ${\tilde \Sigma}_{finite}(p)$ to all orders in perturbation
theory.

{\bf Acknowledgement}
 I am very grateful to Prof. A. Polyakov for his constant
encouragement and
extensive support over the last year and a half and for suggesting
that we
address the quantum stability of the phase transition both in the
model of
rigid QED and that of rigid strings coupled to long range
interactions [6].
I am also grateful to Prof. Y. Nambu for his extensive support,
encouragement
and long discussions over the last two years without which we could
not have
gone far in our investigations. I also thank P.Ramond, C. Thorn, and
Z. Qui
for constructive discussions and suggestions. Finally my gratitude to
my
friend D. Zoller for a fruitful and devoted collaboration over the
years.

{\bf References}

\item {[1]} A. Polyakov, Nucl. Phys. B268 (1986) 406
; A. Polyakov,  Gauge fields, and Strings,
Vol.3, harwood academic publishers
\item {[2]} M. Awada and D. Zoller, Phys.Lett B325 (1994) 115
\item {[3]} M. Awada, D. Zoller, and J. Clark, Cincinnati preprint
June -1 -(1994)
\item {[4]} M. Kalb, and P. Ramond
Phys. Rev. D Vol.9 (1974) 2237
\item {[5]} M. Awada and D. Zoller, Phys.Lett B325 (1994) 119
\item {[6]} A. Polyakov,  Gauge fields, and Strings,
Vol.3, harwood academic publishers, J.Orloff and R.Brout, Nucl.
Phys. B270 [FS16],273 (1986), M. Campostrini and P.Rossi, Phys.
Rev.D 45, 618 (1992) ; 46, 2741 (1992), H. Flyvberg, Nucl. Phys.
B 348, 714, (1991).

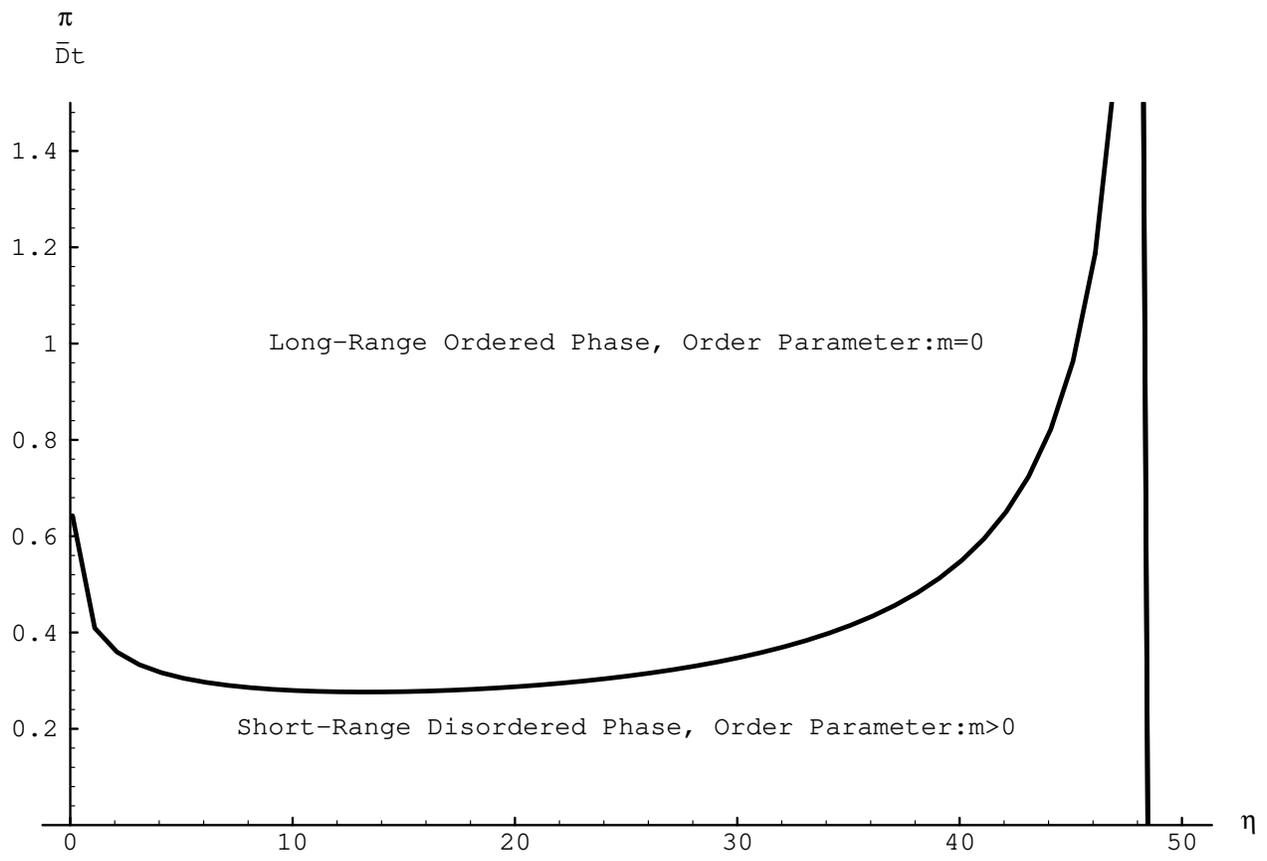
\end